\documentclass[conference,10pt,a4paper]{IEEEtran}
\usepackage[cmex10]{amsmath}
\usepackage{amsfonts,bm}
\usepackage{algorithmic,array,cite,mdwmath,mdwtab,eqparbox}
\ifCLASSINFOpdf
  \usepackage[pdftex]{graphicx}
  \graphicspath{{figures/pdf/}{figures/png/}{figures/jpg/}}
  \DeclareGraphicsExtensions{.pdf,.jng,.jpeg}
\else
  \usepackage[dvips]{graphicx}
  \graphicspath{{figures/eps2/}}
  \DeclareGraphicsExtensions{.eps}
\fi
\usepackage[caption=false,font=footnotesize]{subfig}
\usepackage{fixltx2e}
\usepackage{url}
\usepackage{tikz}
\newcounter{exNo}
\def\vec#1{{\bf#1}}

\def\ket#1{| #1 \rangle}
\def\bra#1{\langle #1 |}

\def\diag{\operatorname{diag}}

\def\H{\mathcal{H}}

\def\Sgn{\mathop{\rm Sgn}}

\begin{document}
\title{Characterization and Control of \\ Quantum Spin Chains and Rings}

\author{\IEEEauthorblockN{Sophie G Schirmer\IEEEauthorrefmark{1},
Frank C Langbein\IEEEauthorrefmark{2} 
}
\IEEEauthorblockA{\IEEEauthorrefmark{1} 
     College of Science (Physics), Swansea University,\\
     Singleton Park, Swansea, SA2 8PP, UK;
     Email: sgs29@swan.ac.uk}
\IEEEauthorblockA{\IEEEauthorrefmark{2}
     School of Computer Science and Informatics, Cardiff University,\\
     5 The Parade, Cardiff, CF24 3AA, UK;
     Email: F.C.Langbein@cs.cardiff.ac.uk}
}
\IEEEspecialpapernotice{(Invited Paper)}
\maketitle
\thispagestyle{empty}
\begin{abstract}
Information flow in quantum spin networks is considered.  Two types of
control -- temporal bang-bang switching control and control by varying
spatial degrees of freedom -- are explored and shown to be effective in
speeding up information transfer and increasing transfer fidelities.
The control is model-based and therefore relies on accurate knowledge of
the system parameters.  An efficient protocol for simultaneous
identification of the coupling strength and the exact number of spins in
a chain is presented.
\end{abstract}

\section{Introduction}
% no \IEEEPARstart

Nature, at a fundamental level, is governed by the laws of quantum
mechanics.  Until recently quantum phenomena were mostly studied by
physicists but significant advances in theory and technology are
increasingly pushing quantum phenomena into the realm of engineering, as
building blocks for novel technologies and applications from chemistry
to computing.  Among the interesting applications are spin networks.
The latter have many potential applications including spintronics and as
networks for transmitting quantum information between processing nodes
on a chip, for example.  It is the latter application that is considered
in this paper.  Information in quantum spin networks is encoded in
quantum states and its propagation is governed by the Schrodinger
equation.  Recent work has shown that this leads to new phenomena such
as the emergence of anti-gravity centers for spin chains \cite{CCA2011}.

The dynamic behavior of even simple spin networks is complicated.  For
example, an excitation created at one end of a linear chain of spins,
does not propagate in a classical fashion, hopping between neighboring
spins from one end to the next.  Rather the excitation creates a
wavepacket, which is a superposition of eigenstates of the Hamiltonian,
that evolves, dispersing and refocusing.  Information transfer from one
spin to another in a spin network is therefore not straightforward.
Neglecting environmental decoherence, quantum transport is determined by
the Hamiltonian of a system $H$ via the Schrodinger equation
\begin{equation}
  \imath\hbar\frac{\partial}{\partial t}\Psi(\vec{x},t) = H \Psi(\vec{x},t),
\end{equation}
where $\imath=\sqrt{-1}$ and $\hbar$ is the Planck constant divided by
$2\pi$.  We shall choose units such that $\hbar=1$.  Expressing
$\Psi(\vec{x},0)$ as a linear combination of eigenfunctions
$\phi_n(\vec{x})$ of the Hamiltonian,
\begin{equation}
  \Psi(\vec{x},0) = \sum_n c_n \phi_n(\vec{x}),
\end{equation}
where $H\phi_n(\vec{x})=E_n\phi_n(\vec{x})$ and $E_n$ is a real number 
corresponding to the energy of $\phi_n(\vec{x})$, we see immediately that
for a static Hamiltonian $H$
\begin{equation}
  \Psi(\vec{x},t) = e^{-\imath H t} \Psi(\vec{x},0) 
                  = \sum_n c_n e^{-\imath E_n t} \phi_n(\vec{x}).
\end{equation}
$\Psi(\vec{x},t)$ governs the evolution of quantum states and the
propagation of information encoded in it.

\section{Quantum Spin Networks}

A quantum spin network for our purposes is simply a collection of $N$
spin$-\tfrac{1}{2}$ particles arranged in space with some of coupling
between spins specified by an interaction Hamiltonian
\begin{equation}
  \label{eq:H}
 H = \sum_{m,n=1}^{N} J_{mn}
      \left(\sigma^x_m \sigma^x_{n} + \sigma^y_m \sigma^y_{n}
                             + \epsilon\sigma^z_m\sigma^z_{n} \right).
\end{equation}
$\epsilon$ is a constant that depends on the type of interaction with
$\epsilon=0$ for XX coupling and $\epsilon=1$ for Heisenberg coupling,
being common.  $J_{mn}$ is the strength of the coupling between spin $m$
and spin $n$, usually proportional to the cubic power of the physical
distance between the two spins.  The factors $\sigma^{x,y,z}_i$ denote
the single spin Pauli operators
\begin{equation*}
\sigma^x= \begin{pmatrix} 0 & 1 \\ 1 & 0 \end{pmatrix}, \quad
\sigma^y= \begin{pmatrix} 0 & -\imath \\ \imath & 0  \end{pmatrix}, \quad
\sigma^z= \begin{pmatrix} 1 & 0  \\ 0 & -1 \end{pmatrix}.
\end{equation*}
$n$ indicates the position of the spin the operator is acting on
\begin{equation*}
\sigma^{x,y,z}_n =
 I_{2\times 2} \otimes \ldots \otimes I_{2 \times 2} \otimes
 \sigma^{x,y,z}\otimes I_{2\times 2}\otimes \ldots \otimes I_{2 \times 2},
\end{equation*}
where the factor $\sigma^{x,y,z}$ occupies the $n$th position among the
$N$ factors.  The system Hilbert space on which $H$ acts is conveniently
taken as $\mathcal{H}:=\mathbb{C}^{2^N}$.

We restrict our attention in this paper to the single excitation
subspace, i.e., it is assumed that the total number of excitations in
the network is one.  The state space of the network is then spanned by
the subset of $N$ single excitation quantum states $\{\ket{n}:
n=1,\dots,N\}$, where $\ket{n}=\ket{\uparrow \uparrow \ldots \uparrow
\downarrow \uparrow \ldots \uparrow \uparrow}$ indicates that the
excitation is localized at spin $n$.  However, unlike in a classical
network the system can be in any \emph{superposition} of these basis
states.  The natural coupling among the spins allows an excitation at
$n$ to drift towards an excitation at $m$, but the {\it fidelity} is
limited by
\begin{equation}
\label{eq:ITC}
\begin{split}
 p_{mn}(t) &= \left| \bra{m} e^{-\imath H_1 t} \ket{n} \right|^2 \\
           &=  \left| \sum_{k=0}^{\tilde{N}} \bra{m}\Pi_k\ket{n} e^{-\imath\lambda_k t}\right| ^2\\
           &\leq \left(\sum_{k=0}^{\tilde{N}} \left| \bra{m}\Pi_k\ket{n} \right|\right)^2=:p_{mn}^*, 
\end{split}
\end{equation}
where $\Pi_k$ is the projector onto the $k$th eigenspace of the
Hamiltonian $H = \sum_k \lambda_k \Pi_k$ and the $\Pi_k, k=1,\dots, N,$
correspond to the single excitation subspace $\H_1$. $p_{mn}^*$,
also referred to as {\it Information Transfer Capacity (ITC)},
is an upper bound on $p_{mn}(t)$, which is attainable if there exist
a time $t\ge 0$ such that
\begin{equation} 
  \label{eq:attain}
   e^{-\imath\lambda_k t} = s_k e^{i\phi} \qquad \forall k \mbox{ s.t. } s_k \neq 0,
\end{equation}
where $s_k = \Sgn(\bra{i} \Pi_k \ket{j}) \in \{0,\pm 1\}$ is a sign
factor and $\phi$ an arbitrary global phase factor.  Terms with $s_k=0$
correspond to eigenspaces that have no overlap with the initial or final
state and can be ignored.  Restricting ourselves to the set $K'$ of
indices for which $s_k=\pm 1$
\begin{equation}
  s_k = \exp\left[-\imath\pi \left(2n_k + \tfrac{1}{2}(s_k-1)\right) \right]
  \qquad \forall k\in K'
\end{equation}
where $n_k$ is an integer.  Inserting this into (\ref{eq:attain}), taking the
logarithm and dividing by $-\imath$
\begin{equation} 
\label{eq2}
   \lambda_k t = 2\pi n_k + \tfrac{\pi}{2}(s_k-1)
   -\phi \qquad \forall k \in K'.
\end{equation}
To obtain meaningful constraints independent of the arbitrary phase
$\phi$ we subtract the equations
\begin{equation}
\label{eq3}
   (\lambda_k-\lambda_\ell) t =  2\pi (n_k-n_\ell) + \tfrac{\pi}{2}(s_k-s_\ell)
  \qquad \forall k,\ell \in K'.
\end{equation}
These conditions are necessary and sufficient for attainability and
physical, involving only differences of the eigenvalues, which are
observable and independent of arbitrary phases.  Broadly speaking the
bounds are attainable in principle, i.e., we can get arbitrarily close to
achieving the maximum transition probability simply by waiting the right
amount of time $t$, if the transition frequencies $\omega_{1k}$ are not
rationally dependent.

In the context of spin chains an attainable upper bound of $1$ for the
transition probability $m\to n$ means that the network is capable for
transferring an excitation from node $m$ to node $n$ with fidelity
arbitrarily close to $1$.  Previous work has shown that the maximum
information transfer capacity between the end spins in a chain with
uniform coupling between nearest neighbors is typically $1$ regardless
of the length of the chain.  However, unlike in the classical case, the
maximum transfer capacity from an end spin to a spin in the middle of
the chain is far less than $1$.  Thus, it is possible for an excitation
to move from one end of the chain to the other without passing through
the center.

\section{Control of Information Transfer}

The information transfer capacities of spin networks are interesting
from a theoretical point of view in terms of understanding information
flow in quantum networks and the restrictions resulting from quantum
mechanical evolution.  However, even if perfect state transfer between
two nodes in a spin network is possible, in practice the time required
to achieve a sufficiently high fidelity can be very long.  For example,
for a spin chain with only five nodes and uniform Heisenberg coupling,
computation of the maximum information transfer capacity and
attainability considerations show that state transfer from $1$ to $5$ can
be achieved with arbitrarily high fidelity.  However, when the transfer
time is limited to say $1000$ time steps in units of $J^{-1}$ then the
attainable fidelity is only around 90\%.  This is where control becomes
relevant.

\begin{figure}
  \includegraphics[width=0.49\textwidth]{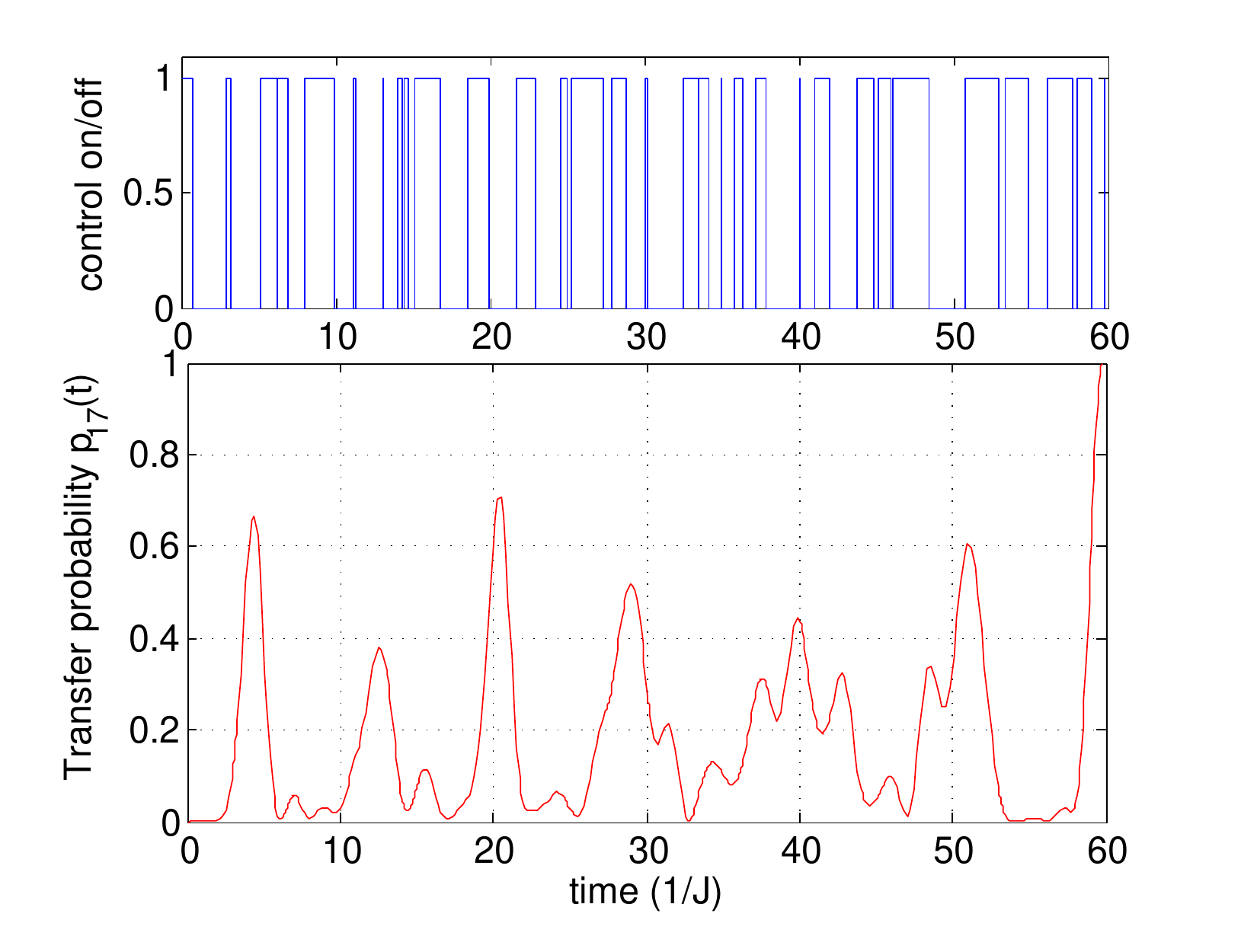}
  \caption{Example of switching control and corresponding evolution of
  transfer probability $p_{17}(t)$ for Heisenberg spin chain of length $7$.}
  \label{fig:switch1}
\end{figure}

Previous work has shown that even bang-bang switching control of a local
perturbation can significantly speed up information transfer between two
ends of a chain with Heisenberg coupling~\cite{PRA2009}.  The type of
control employed in this work was very simple, requiring no more than
bang-bang switching of a fixed \emph{local} perturbation $H_C$ to the
Hamiltonian so that the evolution of the system was governed by the
original system Hamiltonian $H_0$ when the control was turned off, and
by the perturbed Hamiltonian $H_0+H_C$ when the control was turned on.
The control perturbations $H_C$ considered in \cite{PRA2009} involved
local perturbation of the \emph{coupling} $J_{12}$ between the first
pair of spins.  As manipulating the coupling strength between spins can
be difficult to achieve we can alternatively consider a perturbation that
involves local detuning.  Lie algebra considerations show that for a
chain with Heisenberg coupling a simple detuning perturbation of the
form $H_C=\sigma_1^z$ is sufficient for full controllability of the
system on the single excitation subspace~\cite{IEEE2012}.  Optimizing
the switching times as described in~\cite{PRA2009} shows that we can
achieve high-fidelity state transfer.  Fig.~\ref{fig:switch1} shows an
example of an optimal switching sequence and the corresponding transfer
probability $p_{17}(t)$ as a function of time.  We see that we can
indeed achieve constructive interference or refocusing of the excitation
at the target time at the desired end node.  For the purposes of better
illustration of the dynamical evolution we have chosen a short chain but
the procedure is effective for chains with hundreds of nodes, although
the resulting trajectories are extremely complex.

A perhaps even more interesting example for control is information
transfer between nodes in a network beyond what is possible without
control according to the maximum information transfer capacity.  A
simple example of this type is information transfer between nodes in a
ring.  For this arrangement of spins it can be shown that the transfer
fidelity between nodes is limited and usually strictly less than $1$.  In
particular this is the case for information transfer from node $1$ to
$4$, for example, in a ring with seven nodes.  In this case we can
derive analytic expressions for the maximum information transfer
capacity, which show that $p_{14}(t)$ is strictly less than
$1$~\cite{ISCCSP12}.  Unfortunately, unlike for chains, permutation
symmetry of the nodes in a ring shows that applying a control of type
$\sigma_1^{z}$ to the first spin is not sufficient for
controllability.  For example, the operation of swapping spins $k$ and
$N+2-k$ commutes both with $H_0$ and $H_C$ and is therefore a symmetry.
There are two further symmetries and the Lie algebra dimension is only
17, and numerical test of the optimization show that the attainable
fidelity is limited, consistent with the dynamical constraints imposed 
by the symmetries and dark states~\cite{PRA2010}.

\begin{figure}
  \includegraphics[width=0.49\textwidth]{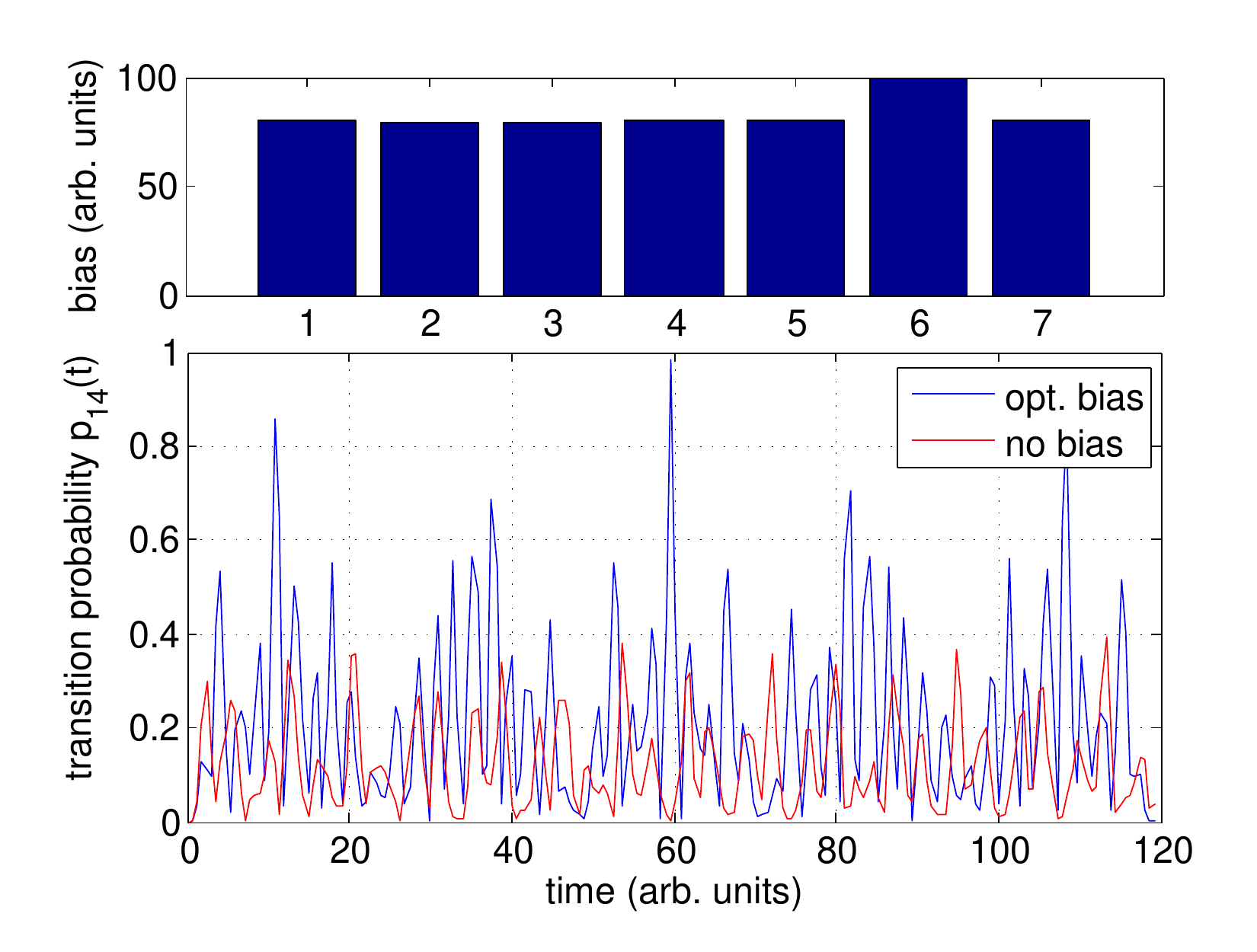}
  \includegraphics[width=0.49\textwidth]{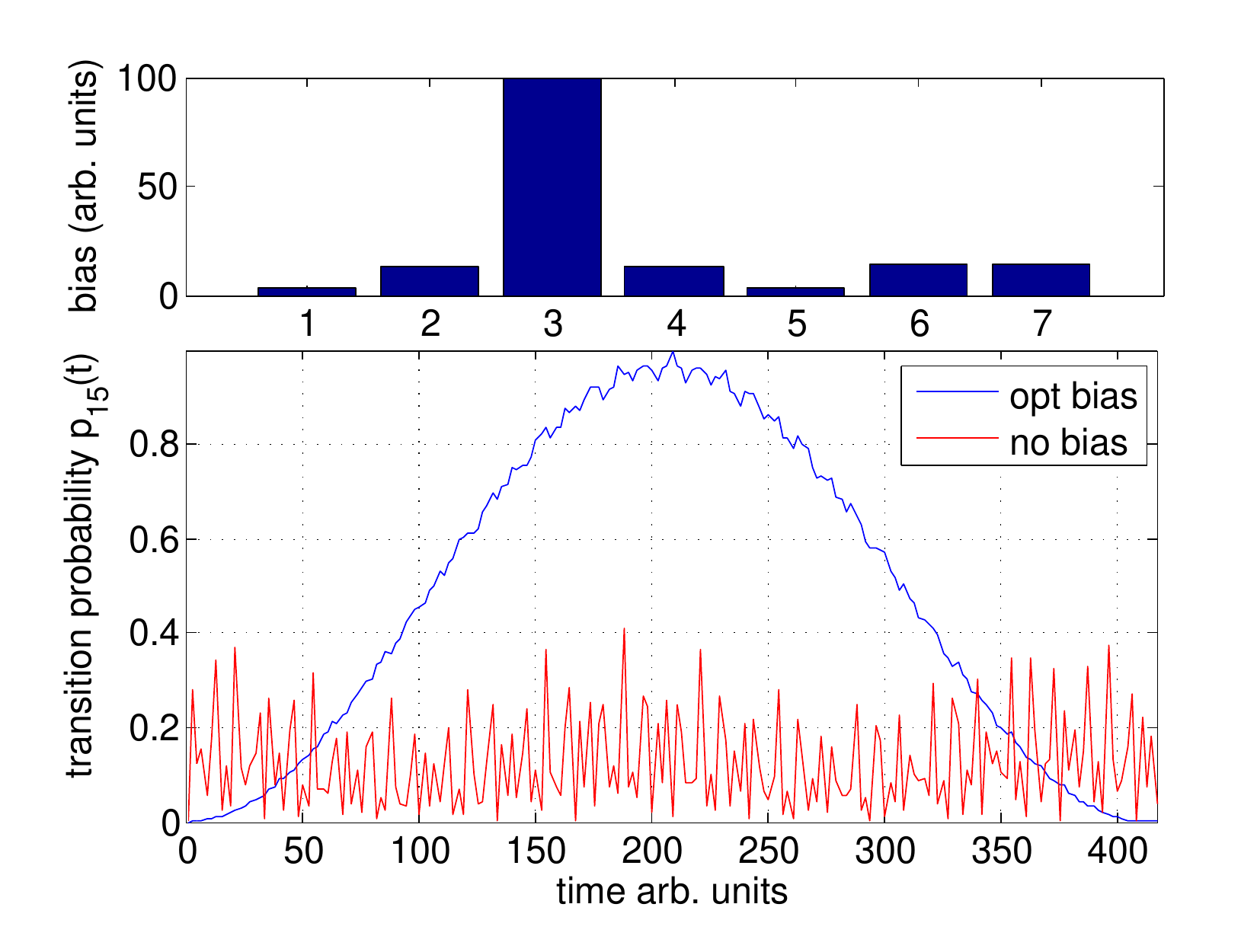}
  \caption{Example of spatial bias control and corresponding evolution of
  transfer probability $p_{14}(t)$ (top) and $p_{15}(t)$ (bottom)
  for Heisenberg spin ring of length $7$.  The red lines in the graph
  show the corresponding transfer probabilities without control.}
  \label{fig:bias1}
\end{figure}

An alternative to temporal on-off switching control is the application
of spatially distributed static perturbations.  In this case we use the
spatial degrees of freedom to control the flow of information.  We
choose the control Hamiltonian to be $H=\diag(c_1,\ldots,c_N)$ and
optimize the spatial biases $c_n$ to maximize the transfer fidelity from
the initial node to the target node at a certain target time.  We can choose
a fixed target time $T$ and find $\vec{c}=(c_n)$ to maximize
\begin{equation}
  |\bra{m} \exp[-\imath T (H_0+H_C(\vec{c}))] \ket{n}|^2, 
\end{equation}
or we can let $T$ vary within a certain range.  An example of the
resulting biases and evolution of the transfer probability for transfer
from node $1$ to node $4$ and $5$, respectively, is shown in
Fig.~\ref{fig:bias1}.  The transfer fidelities are on the order of
$1-10^{-4}$.  The solutions are not unique.  The top graph shows
larger biases with less variation across nodes, which may be preferable
from a practical point of view but has the disadvantage of resulting in
faster oscillations. They are therefore less robust with regard to
variations in the readout time.  The bottom graph shows another solution
with a large bias on one node, which suppresses rapid oscillations in the
transfer probability $p_{15}(t)$ and suggests greater robustness with
regard to readout mistiming. 

\section{Identification of System Parameters}

The examples show that both types of control can be very effective with
suitable optimization.  However, the control design in both cases is
model-based, and in practice the model parameters for spin networks will
often at best be known approximately and may vary, in particular for
engineered systems subject to fabrication tolerances.  Previous work on
control suggests reasonable robustness of optimal controls with regard
to model uncertainties \cite{PRA2009}, but experimental system
identification is expected to be crucial for the success of quantum control
of engineered spin networks.  In practice there are numerous
uncertainties ranging from the network topology, to the precise number
of spins in the network and the strength of the coupling between
connected nodes.

System identification for quantum systems, including spin networks, has
rapidly become a hot topic generating numerous papers
\cite{PRA69n050306(R)} --- \cite{NJP11n103019}.  However, these papers
make numerous assumptions on the available resources, e.g., requiring
prior knowledge of the precise topology, including the exact number of
spins and high-precision quantum state tomography, resources and
knowledge that may not be available in practice.  In the following we
consider a simple protocol to identify both the number of spins in the
network and the coupling strength for rings with uniform coupling using
a simple measurement protocol that requires only a small number of
binary-outcome measurements on fixed spin in the network.

We assume uniform coupling strength and that we can initialize the
system in state $\ket{1}$ and measure it in the same state $\ket{1}$ at
time $t \in [0,T]$, resulting in a measurement outcome of $0$ or $1$
only. $T$ must be sufficiently large to capture the complete dynamics of
this measurement trace. The Hamiltonian of the system is parametrized as
an $N \times N$ matrix
\begin{equation}
 H_1(N,J) = \begin{pmatrix}
0 & J & \ldots & 0 & 0 & 0 & \ldots & 0 & J \\
J & 0 & \ldots & 0 & 0 & 0 & \ldots & 0 & 0 \\
\vdots & \vdots & \ddots & \vdots & \vdots & \vdots & & \vdots & \vdots \\
0 & 0 & \ldots & 0 & J & 0 & \ldots & 0 & 0 \\
0 & 0 & \ldots & J & 0 & J & \ldots & 0 & 0 \\
0 & 0 & \ldots & 0 & J & 0  & \ldots & 0 & 0\\
\vdots   & \vdots & & \vdots & \vdots & \vdots & \ddots & \vdots & \vdots \\
0 & 0 & \ldots & 0 & 0 & 0 & \ldots & 0 & J \\
J & 0 & \ldots & 0 & 0 & 0 & \ldots & J & 0
\end{pmatrix}
\end{equation}
The probability of measuring $1$ at time $t$ is
\begin{equation}
  \theta(N,J,t) = |\bra{1}e^{-iH_1(N,J)t}\ket{1}|^2.
\end{equation}
%F Insert he eigenvalues here? No space (4 page limit)
So given $M$ measurement results $E$ for times $t_1,\dots,t_m$, each
repeated $R_m$ times, the probability of $H_1(N,J)$ being the correct
Hamiltonian is given by the binomial distribution
\begin{gather} 
P(H_1(N,J)|E)  = \nonumber\\
   \prod_m {A_m \choose R_m} \theta(N,J,t_m)^{A_m}(1-\theta(N,J,t_m)(t_m))^{R_m-A_m}
\end{gather}
where $A_m$ is the number of $1$ measurements at time $t_m$. For
numerical reasons use instead the log likelihood
\begin{equation}
L(N,J|E) = -\log P(H_1(N,J|E)).
\end{equation}

To estimate the parameters $N$ and $J$, we iteratively take $M$
measurements at times $t_m$ sampled in the time interval $[0,T]$ using
the Hammersley low-discrepancy sequence. We then sample $L$ over a given
parameter domain $\{N_{\min}, N_{\min}+1, \dots, N_{\max}\} \times
[J_{\min},J_{\max}]$ with $K$ sample points per ring-size parameter $N$.
Initially $K$ points $p_k$ are sampled uniformly for each size $N$ in
the coupling strength parameter interval $[J_{\min},J_{\max}]$.  $L$ is
evaluated at each sample point $p_k$.  Then the $K$ samples are resampled
according to the sampling density function
\begin{equation}
D(j) = \frac{1}{2} (p_{k+1} - p_{k-1}) L(N, j|E).
\end{equation}
where $k$ is chosen such that $p_k$ is closest to $j$, $p_{-1} = p_0 =
0$ and $p_{K+1} = p_K = T$ at the interval boundaries.  Iteratively
additional measurements are taken by continuing the Hammersley sequence
on the measurement trace, updating the $L$ values at the sample points
to update the log-likelihood for the new measurements, and then resample
the coupling strength intervals.

This is repeated until the sampled log-likelihood over the parameter
domain has a clear peak according to the sample points taken. Then the
$N$ value with the largest $L$ value for the sampled points is selected
and a simple 1D maximization strategy, such as hill-climbing is used to
find the coupling strength for which the log-likelihood is largest.

The samples per coupling strength interval $K$ can typically be small,
say about $50$ as the resampling process moves them towards the highest
peaks in the interval and tracks these peaks. It is typically also
sufficient to only take a few measurements $M$, say $10$ per iteration,
with a low number of repetitions $R$, say $10$. After a few iterations
(about $10$), the log-likelihood has a clear peak and ring size as well
as coupling strength can then easily be estimated to high accuracy.

\begin{figure}
\includegraphics[width=0.49\textwidth]{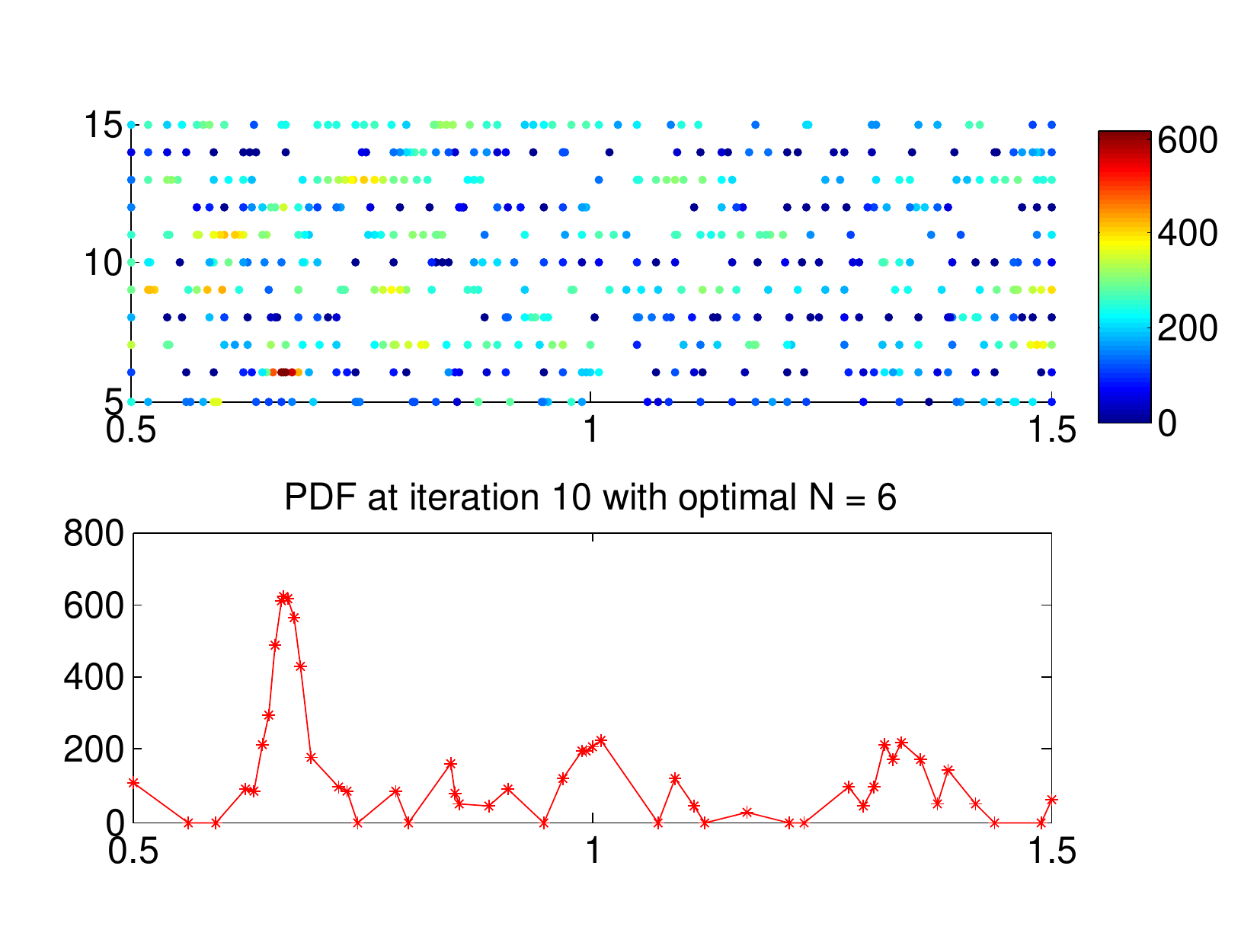}
\caption{Network identification example for a ring with $6$ nodes and
coupling strength $0.666$. Top: log-likelihood sample positions in the
parameter domain $\{5,6,\dots,15\} \times [.5,1.5]$, after $10$
iterations of taking $10$ time samples, repeated $10$ times, showing a
clear peak close to the exact parameter values. Bottom: log-likelihood
function over the coupling strength parameter domain $[.5,1.5]$ for
$N=6$.}  \label{fig:ident1}
\end{figure}

Fig.~\ref{fig:ident1} shows an example for a ring with $6$ spins and
coupling strength $0.666$. After $10$ iterations, adding new sample
times for the measurements, each measured $10$ times, and then the final
optimization for the coupling strength resulted in estimating the ring
size clearly to be $6$ with a coupling strength of $0.666083$, i.e. an
error of $8.3 \times 10^5$.

\section{Conclusion}

We have demonstrated a bang-bang and a bias control scheme for
optimizing the information transfer between nodes in simple quantum spin
networks, either ensuring maximal transfer in limited time or enabling
maximal transfer against the natural dynamics of the network.  We have
further presented a scheme to simultaneously estimate the network size
and coupling strength for simple networks with uniform coupling using an
efficient sampling strategy to take measurements as well as sampling the
parameter domain.  Future work will extend the control and
characterization schemes to more general and realistic quantum networks.

\end{document}